\documentstyle[preprint,floats,epsf,aps]{revtex}

\begin{document}
\draft


\title {\bf Accurate construction of transition metal pseudopotentials}
\author{Ilya Grinberg, Nicholas J. Ramer,\cite{NJR} and Andrew M. Rappe}
\address{Department of Chemistry and Laboratory for Research on the
Structure of Matter\\ University of Pennsylvania, Philadelphia, PA
19104-6323} 
\maketitle

\begin{abstract}
We generate a series of pseudopotentials to examine the relationship between pseudoatomic properties and solid-state results. We find that lattice constants and bulk moduli are quite sensitive to eigenvalue, total-energy difference and tail norm  errors, and clear correlations emerge. These trends motivate our identification of two criteria for accurate transition metal pseudopotentials. We find that both the preservation of all-electron derivative of tail norm with respect to occupation and the preservation of all-electron derivative of eigenvalue with respect to occupation {[Phys. Rev. B {\bf 48}, 5031 (1993)]} are necessary to give accurate bulk metal lattice constants and bulk moduli.  We also show how the fairly wide range of lattice constant and bulk modulus results found in the literature can be easily explained by pseudopotential effects.

\end{abstract} 

{\em Ab} {\em initio} density functional theory~\cite{HK,KS} (DFT) calculations have been widely used over the past ten years to study transition metal surfaces and surface-adsorbate systems. The plane-wave pseudopotential (PSP) method accounts for most of these DFT calculations due to its superior speed, which allows one to study computationally intensive problems inaccessible with other methods.  The absence of core electrons in the calculation and the reduced-cutoff plane-wave expansion of the PSP greatly reduce the computational cost of the solid-state calculation.  Even though the PSP is required to agree with the all-electron potential outside a specified core radius ($r_c$), the mapping of a real, physical system onto an artificial one constructed of PSPs can introduce errors in the calculation. An ideal PSP should be completely transferable, i.e. it will mimic perfectly the behavior of the all-electron nucleus and core potential in various local chemical environments and will produce solid-state and molecular  results identical to those of an all-electron calculation.  Methods capable of generating transferable PSPs with small plane-wave cutoffs have been developed over the past twenty years.\cite{HSC,KB,RappePS,RRPS,Vdblt,GTH,Vackar}.  

It is  widely assumed  that the magnitude of the pseudopotential error is less than  that of other approximations used in PSP-DFT calculations.  Nevertheless, some fundamental questions regarding PSP construction remain unresolved. While it is accepted that a PSP must preserve certain all-electron properties to be considered transferable, it is unclear which all-electron properties are vital.  

Various criteria for comparing PSP results to those of an  all-electron potential have been proposed, such as agreement between all-electron and PSP eigenvalues along with total-energy differences, norm-conservation at the reference configuration and preservation of logarithmic derivatives at $r_c$,~\cite{HSC} and the correct chemical hardness matrix.~\cite{Teter} Eigenvalue and total-energy difference  agreement are the most often used criteria for determining if a PSP is transferable. However, no clear correlation has been firmly established between these criteria and solid-state results. The determination of PSP quality is further confused by the fact that a simple comparison with experiment sometimes  cannot be used to gauge PSP quality.  For example, since the local-density approximation (LDA) overbinds, an LDA-PSP calculation that predicts the experimental value for the lattice constant is less accurate than an LDA-PSP calculation that slightly underestimates the bulk lattice constant.  

Bulk lattice constants and moduli are often used to assess the quality of a PSP. These parameters can now be easily determined due to powerful computers and fast DFT packages.  PSP construction is also a fairly routine procedure due to the availability of various efficient PSP generation codes. The PSP transferability error is widely considered to be insignificant compared to  other approximations used in PSP-DFT calculations, such as the choice of the exchange-correlation functional.  Yet, surveying the {\em ab} {\em initio} PSP-DFT calculations performed in the past five years on various transition metals, transition metal surfaces and adsorbate systems, one finds a broad distribution in predicted lattice constants and bulk moduli for both LDA and generalized gradient approximation (GGA) calculations~\cite{Eichler,Hafner,EH,Feibelman,Bohnen,Cho,Stampfl,GPS,Schwegmann,Norskov,Hammer,Mavrikakis,Stockbro}, with variations in results being greater than that caused by the use of different exchange-correlation functionals. Lattice constant errors in published LDA calculations vary from -1.3$\%$~\cite{Stockbro} to +0.3$\%$~\cite{Cho} to +1.6$\%$~\cite{Stockbro}  for rhodium  and from -1.0$\%$~\cite{Ratsch} to -0.7$\%$~\cite{Baroni} to +1.7$\%$~\cite{Yu} for silver.   For bulk ruthenium metal, GGA lattice constants with errors of +1.5$\%$~\cite{HammerS}, +2.0$\%$~\cite{Stampfl} and +3.0$\%$~\cite{Schwegmann} have been calculated.  For a wide range of materials, it is known that LDA underestimates bond lengths somewhat, while GGA slightly overestimates bond lengths.  Nevertheless, many PSP-DFT studies on transition metals in the literature use LDA PSPs that give  lattice constants larger than experiment, sometimes by as much as 1$\%$.~\cite{EH,Cho,Stockbro}    While the LDA lattice constant errors in the literature are typically between -1$\%$ to +1$\%$, bulk moduli are overestimated by up to 30$\%$.~\cite{Bohnen,Stockbro} The use of GGA functionals reduces the tendency to overbinding, so the range of GGA error in reported calculations shifts to 0$\%$ to 3$\%$ error in the lattice constant and -10$\%$ to 10$\%$ error in the bulk modulus.~\cite{Bohnen,Hammer} The inability of the otherwise reliable PSP-DFT methods to correctly and consistently predict these simple quantities is puzzling.  Since the methods as well as the exchange-correlation functionals used to obtain these bulk parameters are identical or very similar for all of these studies, much of the variation in the literature must be due to the use of different PSPs. 
	
In order to find the possible sources of PSP error and to test for correlations between these errors and errors in the solid-state results, we must examine the PSP construction procedure.  An ideal PSP should reproduce the all-electron wave functions beyond $r_c$ for all atomic configurations.  Enforcing eigenvalue agreement for all atomic configurations will not necessarily enforce wave function agreement. Although the eigenvalue governs the rate of exponential decay of both the all-electron and pseudo-wave functions, the prefactors multiplying the exponentials may be different. This will lead to either underestimation or overestimation of the charge in the tail of the wave function, and the incorrect distribution of charge on the atomic level will lead to inaccuracies in the solid-state properties. The widely used norm-conserving PSP construction methods enforce agreement of the wave function norm for the reference configuration only. Therefore, norm conservation  and eigenvalue agreement in configurations other than the reference configuration are important for PSP transferability, and both must be checked to gauge the quality of the PSP at the atomic level. This is an extension of the original idea of Hamann, Schl\"uter and Chiang in their landmark paper on norm-conserving PSPs.~\cite {HSC} They showed that enforcing norm conservation at the reference configuration dramatically improved PSP quality.  In this paper, we demonstrate that the failure to conserve norm in configurations other than the reference configuration leads to significant transferability errors, and we propose a new atomic transferability criterion which leads to more accurate atomic and solid-state results.

The designed nonlocal (DNL)  PSP construction approach~\cite{RRPS} allows us to adjust the amount of PSP norm and eigenvalue error in various atomic configurations, while leaving the agreement at the reference configuration unchanged.  We can, therefore,  systematically introduce errors in different proposed transferability criteria to examine the consequences of PSP error in solid-state calculations. We can then  approximately  enforce  these new criteria and determine  whether accuracy tracks quantitatively with criterion enforcement.

  In solids, the interatomic potential can be heuristically described as the sum of an attractive bonding term and  repulsive Pauli and electrostatic interaction terms.  The balance between these terms leads to an equilibrium lattice constant, and a change in either the repulsive or the attractive part of the potential will change the solid-state bulk properties of the crystal.  Increasing the amount of charge in the tail region will strengthen the interatomic repulsive potential, expanding the equilibrium lattice constant of the crystal.  In the same way, decreasing the charge in the tail will contract the crystal.  Errors in the eigenvalues and total-energy differences affect the attractive bonding term of the solid-state interatomic potential.  The total-energy differences between the various atomic configurations are related to $d$$\rightarrow$$s$ excitation energies, which govern the extent of $sd$ hybrid orbital formation in solids.  An overestimation of the  $d$$\rightarrow$$s$ excitation energy implies an increase in the  hybridization energy cost, leading to a weaker bonding term and expansion of the crystal.  Conversely, underestimation of $d$$\rightarrow$$s$ excitation energy will lead to contraction of the crystal.

  To verify  our understanding of  how norm-conservation and eigenvalue agreement affect solid-state results, we examine six different Rh PSPs. The $d$-states are more populated than the $s$-states in the right half of the $d$-block of the Periodic Table and unlike the $s$-states, the $d$-states are localized on individual atoms in the solid state. We therefore will focus on the norm of the $d$-states.  We  also compare eigenvalues and total-energy differences. All calculations are done using the Perdew-Burke-Ernzerhof GGA~\cite{PBE} exchange-correlation functional at a plane-wave cutoff of 50~Ry.   For all six PSPs, we compute eigenvalues, total-energy differences, and norms of the tail region ($r$ $\geq$ 2.6 Bohr) for three sample configurations: $s^0p^0d^9$, $s^1p^0d^8$ and $s^2p^0d^7$.  For comparison, we also compute the corresponding all-electron values in the three sample configurations.  These configurations span the spectrum of neutral Rh states important in the bulk metal solid. PSPs Rh$_{\rm A}$, Rh$_{\rm B}$, Rh$_{\rm C}$, Rh$_{\rm D}$ and Rh$_{\rm E}$ were created in a +0.1 ionized reference configuration ($s^0p^0d^{8.9}$) by gradually varying the depth of the DNL augmentation operator $\widehat{A}$. PSP Rh$_{\rm A}$ preserves all-electron $d$-state norms and eigenvalues and total-energy differences, PSP Rh$_{\rm B}$ and PSP Rh$_{\rm C}$ are constructed to match the all-electron $d$-state norms but not the eigenvalues and total-energy differences, and PSP Rh$_{\rm D}$ is constructed so that the eigenvalues and total-energy differences match the all-electron results, but the  $d$-state norms do not. PSP Rh$_{\rm E}$ is constructed to give a lattice constant error of +2$\%$ typical for GGA calculations found in the literature. PSP Rh$_{\rm F}$ was created in a +1.7 ionized reference configuration ($s^{0.1}p^{0.1}d^{7.1}$) and $\widehat{A}$ was adjusted until solid-state calculations gave a -0.2$\%$ error in the lattice constant.  PSP Rh$_{\rm F}$ has large errors in both eigenvalue and total-energy differences and norms but gives a lattice constant similar to that of  PSP Rh$_{\rm A}$. The results for the six PSP's as well as all-electron values  are in Table I.

Examination of the results for PSPs Rh$_{\rm A}$ and Rh$_{\rm D}$ in Table I shows that underestimation of charge in the tail  region ($r$ $\geq$ 2.6 Bohr)  decreases the solid-state lattice constant. Despite having very similar eigenvalues to PSP Rh$_{\rm A}$,  pseudo-wave functions of PSP Rh$_{\rm D}$ are shifted inward as shown by the negative norm error.   This makes the repulsive term in the interatomic potential smaller, resulting in a smaller lattice constant in the solid state.  The results in Table I also confirm the importance of eigenvalue and total-energy difference agreement and the connection between $d$$\rightarrow$$s$ excitation energies and solid-state lattice constant.  A comparison of PSP Rh$_{\rm A}$ and PSP Rh$_{\rm B}$ shows that an average increase in $d$-eigenvalue error of only 2.0 mRy per configuration and total-energy difference error of about 1.0 mRy per configuration leads to an expansion of the crystal by about 0.8$\%$.   An overestimation of the  $d$$\rightarrow$$s$ excitation energy by PSP Rh$_{\rm B}$ implies an increased  hybridization energy cost.  This means that PSP Rh$_{\rm B}$ will underestimate the attractive bonding energy term, leading to an overestimation of the lattice constant. On the other hand if the hybridization energy is underestimated, the attractive bonding term will be larger and the lattice constant will be smaller.  In the case of PSP  Rh$_{\rm C}$,  average $d$-eigenvalue error of  +5.0 mRy per configuration and average total-energy difference error of about -3.6 mRy per atom leads to a contraction of the crystal, giving an error of -2.3$\%$ in the lattice constant.   Not only the direction but also the magnitude of the lattice constant error for  PSP Rh$_{\rm B}$ and PSP Rh$_{\rm C}$ track with the respective total-energy difference errors.  Since many  GGA calculations in the literature overestimate the lattice constant by 2$\%$, we created PSP Rh$_{\rm E}$ to show that this could be accounted for by pseudopotential effects.  An overestimation of the $d$$\rightarrow$$s$ excitation energy by an average of 3.2 mRy is all that is required to cause  the 2$\%$ expansion in the solid state.

An accurate GGA lattice constant can be obtained even by a PSP with incorrect bonding and repulsion terms, if the two errors cancel.  However, the errors will not cancel out for the bulk modulus.   If interatomic attraction and repulsion  are equally overestimated, the potential well will be steeper, resulting in a significant overestimation of bulk modulus. To examine this effect we can compare the data for PSPs Rh$_{\rm A}$ and Rh$_{\rm F}$. The  lattice constant error is very small for both potentials, which is what we expect from GGA calculations. However PSP Rh$_{\rm F}$ has significant eigenvalue, total energy and   norm transferability errors. The error in the bulk modulus changes considerably from PSP Rh$_{\rm A}$ to PSP Rh$_{\rm F}$. As can be seen from the data, the $d$$\rightarrow$$s$ excitation energy is significantly underestimated, leading to  greater attractive bonding energy and consequently smaller lattice constant. However more charge in the  tail region (shown by increased norms in sample states) leads to a greater repulsive potential and cancels out the effect of the eigenvalue/total-energy difference error. The steeper potential that is produced by the combination of the two effects results in the overestimation of the bulk modulus for PSP Rh$_{\rm F}$.  Superficially, PSP Rh$_{\rm F}$ seems superior to PSP Rh$_{\rm A}$ due to smaller bulk modulus error.  However, the better comparison with experiment is due to a fortuitous cancellation of pseudopotential and GGA functional errors which will not necessarily be favorable in calculations for other solid-state properties.

\begin{table}[t]
\caption{Pseudopotential (PSP) results for Rh.  Total-energy
differences ($\Delta E_{\rm tot}$), $d$-eigenvalues ($\varepsilon_d$)
and $d$-state charge in the tail  region  ($N_d$) are given for an all-electron atom (AE).  Absolute
errors are given for the PSPs described in text. All energies are in
Ry.  Percent errors in lattice constant ($R$) and bulk modulus ($B$)
are given for solid-state calculations using  the PSPs.}

\begin{tabular}{rrrrc}
 &\multicolumn{1}{c}{$\Delta E_{\rm tot}$}&\multicolumn{1}{c}{$\varepsilon_d$}&\multicolumn{1}{c}{$N_d$}&\multicolumn{1}{c}{$R$,$B$ Error}
 \\ 
\hline


Rh AE  $s^0p^0d^9$&   	0.0000& 	-0.2678& 	 0.1328&  \\
  $s^1p^0d^8$&   	0.1121& 	-0.4637& 	 0.0988& \\
  $s^2p^0d^7$&   	-0.3572& 	-0.6878& 	 0.0732&\\

Rh$_{\rm A}$ PSP $s^0p^0d^9$&   	0.0000& 	 0.0008& 	0.0000&  0.0 \\
 $s^1p^0d^8$&   	0.0001& 	-0.0005& 	-0.0002& -12 \\
 $s^2p^0d^7$&   	-0.0003& 	 0.0014& 	-0.0002&\\

Rh$_{\rm B}$ PSP $s^0p^0d^9$&   	 0.0000&	-0.0009& 	 0.0002& 0.8\\
 $s^1p^0d^8$&   	 0.0005&	-0.0019& 	-0.0002& -13\\
 $s^2p^0d^7$&   	 0.0024&	-0.0028& 	-0.0002 &\\

Rh$_{\rm C}$ PSP $s^0p^0d^9$&   	 0.0000&	-0.0006& 	 0.0000&-2.3\\
P $s^1p^0d^8$&   	-0.0017&	 0.0041& 	-0.0001& -13\\
 $s^2p^0d^7$&   	-0.0092&	 0.0117& 	-0.0002&\\

Rh$_{\rm D}$ PSP $s^0p^0d^9$&   	 0.0000&	-0.0014&	 0.0002 &-1.3\\
 $s^1p^0d^8$&   	 0.0000&	 0.0000&	-0.0004&-10\\
 $s^2p^0d^7$&   	-0.0002&	-0.0006&	-0.0009 &\\
Rh$_{\rm E}$ PSP $s^0p^0d^9$&   	 0.0000&	 0.0003&	 0.0003 & 2\\
 $s^1p^0d^8$&   	 0.0010&	 -0.0038&	-0.0001& -15\\
 $s^2p^0d^7$&   	 0.0053&	 -0.0067&	 0.0004 &\\

Rh$_{\rm F}$ PSP $s^0p^0d^9$&   	  0.0000&	 0.0065&	 0.0037 &-0.2\\
 $s^1p^0d^8$&   	 -0.0140&	 0.0051&	 0.0017&  3\\
 $s^2p^0d^7$&   	 -0.0205&       -0.0018&	 0.0006 &\\

\end{tabular}
\end{table}

While the eigenvalue and norm-conservation errors are comprehensive quantities (i.e. sums over various atomic configurations), the derivative of the amount of charge in the tail region in state $i$ ($N_i$) with respect to occupation of state $j$ $\left(\frac {dN_i}{df_j}\right)$ and the derivative of eigenvalue of state $i$ with respect to occupation of state $j$ $\left( \frac{d\varepsilon_i}{df_j}\right)$ are properties of the reference configuration only, making them more amenable to enforcement in the PSP construction. These two tensors are also good predictors of norm-conservation and eigenvalue error in sample configurations.  It has been shown~\cite{Teter} that $\frac{d\varepsilon_i}{df_j}$ (the chemical hardness)  is important for transferability.  

In this work we confirm the importance of $\frac{d\varepsilon_i}{df_j}$  conservation while adding $\frac{dN_i}{df_j}$ as a second transferability criterion.  To show how the quality of PSP $\frac {dN_i}{df_j}$ and $\frac{d\varepsilon_i}{df_j}$ affect solid-state accuracy we create a PSP quality correlation map for Rh (Figure 1) . The PSPs were created in various reference configurations ranging from +1.7 ionized to neutral with various $r_c$'s, plane-wave cut-off energies and augmentation operators.  The abscissa of Figure 1 is the error in $\frac{d\varepsilon_i}{df_j}$ and the ordinate is the error in $\frac {dN_i}{df_j}$.  The quality of each potential is evaluated based on the lattice constant and bulk modulus obtained from solid-state DFT calculations.   It can be seen that the best PSPs (those with small error in $\frac {dN_i}{df_j}$ and $\frac{d\varepsilon_i}{df_j}$) fall close to the origin, while a deviation in either of the atomic quantities strongly degrades solid-state results. This confirms the importance of these two criteria, and therefore both should be included in PSP construction.

 We have found these same trends in calculations using the LDA exchange-correlation functional~\cite{PZ,CA} and in calculations on other transition metals including Ru, Pd, Pt, Cu, Ag and Au. Chemisorption energies of atoms and molecules on transition metal surfaces obtained by PSP-DFT calculations will also be affected by PSP error due to the strong dependence of the chemisorption energy on lattice constant~\cite{Mavrikakis}. 

In this paper, we have presented results for  bulk Rh metal properties calculated with the GGA exchange-correlation functional. We constructed a family of Rh pseudopotentials with various eigenvalue, total-energy difference  and tail norm conservation properties. We then calculated lattice constants and bulk moduli for each pseudopotential to gauge how the atomic level errors correlate with the results of the solid-state calculations. We found that the bulk solid-state properties are very sensitive to the choice of pseudopotential.  The range of the results  given by various PSPs (-2.3$\%$ to +2.0$\%$) is considerably larger than the commonly assumed range of results  given by various exchange-correlation functional approximations.  We find that simultaneously enforcing agreement between all-electron and pseudopotential $\frac {dN_i}{df_j}$ and $\frac{d\varepsilon_i}{df_j}$   greatly reduces Rh pseudopotential error, leading to lattice constant which  is slightly larger than  experiment and a bulk modulus which is somewhat smaller than experiment.   We show that the  simultaneous enforcement of $\frac {dN_i}{df_j}$ and $\frac{d\varepsilon_i}{df_j}$ agreement will give accurate bonding and repulsive forces leading to accurate solid-state properties.

This work was supported by NSF grant DMR 97-02514 and the Office of
Naval Research grant No. N-00014-00-1-0372. AMR acknowledges the
support of the Camille and Henry Dreyfus Foundation. Computational support was
provided by the San Diego Supercomputer Center and the National Center
for Supercomputing Applications.

\newpage
\begin{figure}[p]
\epsfysize=7.5in
\centerline{\epsfbox[0 0 530 530]{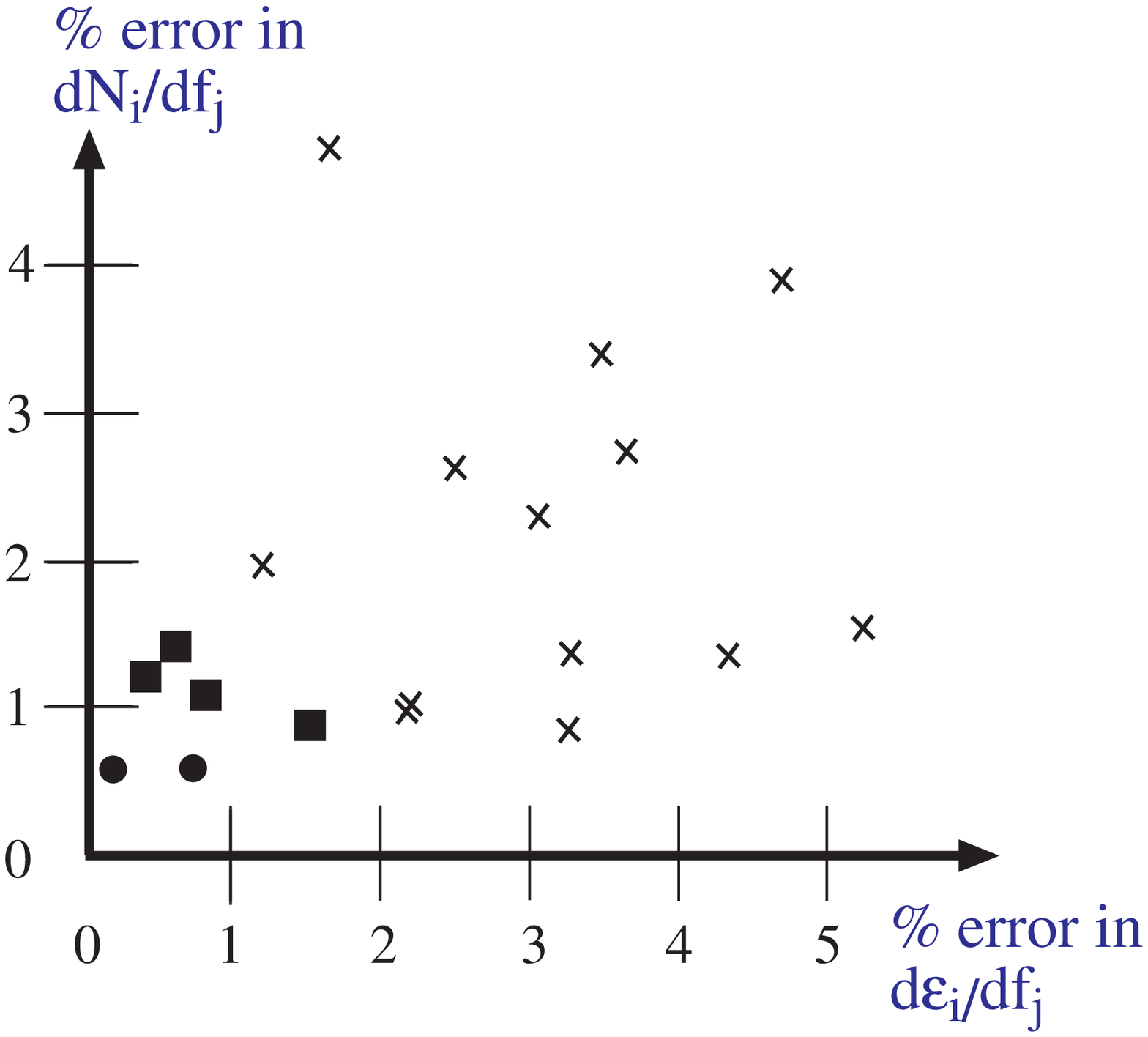}}
\caption{Rh pseudopotential (PSP) quality assessment at the atomic
level.  Circles correspond to lattice constant error of -0.5$\%$ to 0.5$\%$. Squares correspond to lattice constant error of -1.5$\%$ to -0.5$\%$ and +0.5$\%$ to +1.5$\%$. $\boldmath \times$'s correspond to lattice constant error of -3.5$\%$ to -1.5$\%$ and +1.5$\%$ to +3.5$\%$.  A significant deviation from the  bulk modulus obtained for Rh$_{\rm A}$ will change the ranking of a PSP from a circle to a square  or from a square to an $\boldmath \times$. }
\end{figure}
\end{document}